\documentclass[12pt]{article}%  14.11.24  безотражательные решения
\usepackage[english]{babel}
\usepackage[cp1251]{inputenc}%
\usepackage{graphicx}
\usepackage{epsfig}
\usepackage{ulem}
\usepackage[dvipsnames]{color}
\begin{document}
\newcommand{\ba}{\begin{array}}
\newcommand{\ea}{\end{array}}
\newcommand{\be}{\begin{equation}}
\newcommand{\ee}{\end{equation}}
\newcommand{\bea}{\begin{eqnarray}}
\newcommand{\eea}{\end{eqnarray}}
\newcommand{\bfig}{\begin{figure}}
\newcommand{\efig}{\end{figure}}
\newcommand{\Bl}{\Bigl}
\newcommand{\Br}{\Bigr}
\newcommand{\RE}{{\rm Re}\,}
\newcommand{\IM}{{\rm Im}\,}
\newcommand{\re}{{\rm e}}
\newcommand{\cD}{{\cal D}}
\newcommand{\cF}{{\cal F}}
\newcommand{\cH}{{\cal H}}
\newcommand{\bB}{{\bf B}}
\newcommand{\bV}{{\bf V}}
\newcommand{\bb}{{\bf b}}
\newcommand{\bv}{{\bf v}}
\newcommand{\hL}{\hat{L}}
\newcommand{\tF}{\tilde{F}}
\newcommand{\tP}{\tilde{P}}
\newcommand{\tif}{\tilde{f}}
\newcommand{\tig}{\tilde{g}}
\newcommand{\tih}{\tilde{h}}
\newcommand{\ts}{\tilde{s}}
\newcommand{\tu}{\tilde{u}}
\newcommand{\tU}{\tilde{U}}
\newcommand{\tv}{\tilde{v}}
\newcommand{\ptl}{\partial}
\newcommand{\dd}{{\rm d}}
\newcommand{\al}{\alpha}
\newcommand{\gm}{\gamma}
\newcommand{\Gm}{\Gamma}
\newcommand{\Lb}{\Lambda}
\newcommand{\lb}{\lambda}
\newcommand{\vp}{\varphi}
\newcommand{\vep}{\varepsilon}
\newcommand{\bvp}{\bar{\varphi}}
\newcommand{\bP}{\bar{\Phi}}
\newcommand{\om}{\omega}
\newcommand{\Om}{\Omega}
\newcommand{\sg}{\sigma}
\newcommand{\din}{\displaystyle\int\limits}
\newcommand{\IN}{\displaystyle\int\limits_{0}^}
\newcommand{\II}{\displaystyle\int\limits^{\infty}_}
\newcommand{\IE}{\displaystyle\int\limits^{1}_}
\newcommand{\dfrac}{\displaystyle\frac}
\renewcommand{\abstractname}{}
 \title{On the solutions of a factorized wave equation}
 \author{Semyon Churilov,\\
 {\footnotesize\it Institute of Solar-Terrestrial Physics of the Siberian
 Branch of Russian Academy of Sciences,}\\
 {\footnotesize\it 126a Lermontov St., Irkutsk, 664033, Russia.}}%
 \date{}
 \maketitle
 \vspace{-10mm}
 \begin{abstract}\noindent
 Long-distance transmission of energy by waves is a key mechanism for many
 natural processes. It becomes possible when the inhomogeneous medium is
 arranged in such a manner that it enables a specific type of waves to
 propagate with virtually no reflection or scattering. If the corresponding
 wave equation admits factorization, at least one of the waves it describes
 propagates without reflection. The paper is devoted to searching for
 conditions under which both solutions of a one-dimensional factorized wave
 equation of the second order describe traveling waves, that is, waves
 propagating without reflection. Possible variants of wave structure are
 found and the results are compared with those obtained in previous studies.
 \end{abstract}
 \section{Introduction}
 \label{sec1}
 \hspace\parindent
 Studying the possibility of wave propagation without reflection and
 scattering in inhomogeneous environment is necessary to explain phenomena
 of energy transmission by waves over long distances observed in nature. In
 the Earth's atmosphere, the energy is transferred from the lower layers to
 the upper ones by acoustic-gravity and internal waves \cite{Grig,Achatz}.
 In the seas and oceans, storm waves and tsunamis are able to transmit the
 energy over long distances \cite{Vallis,Levin}. In space and astrophysical
 plasmas, this work is often performed by Alfv\'{e}n waves \cite{Alfven},
 that is, by transverse (with respect to the directions of the magnetic
 field and phase velocity) oscillations which practically do not disturb the
 density. In particular, it is Alfv\'{e}n waves that play a key role in the
 transfer of energy from the lower solar atmosphere into the corona and in
 the corona heating (see, for example, \cite{Arregui,Moortel} and references
 cited).

 Theoretical search for configurations of an inhomogeneous medium that allow
 wave propagation without reflection is very difficult and analytical
 results have been obtained, as a rule, for cases of one spatial dimension
 (or those reducible to one-dimensional) which we shall assume below. The
 main way to finding the parameters of inhomogeneous non-reflective media is
 in reducing the equation for waves to some reference equation. For media at
 rest, the equations with coordinate-independent coefficients are usually
 chosen as such, for example, the classical equation for waves propagating
 with a velocity $c$,
 \be
 \dfrac{\ptl^2 f}{\ptl t^2} - c^2\dfrac{\ptl^2 f}{\ptl x^2} = 0,
 \label{WE}
 \ee
 whose general solution
 \[
 f(x,t) = F_1\left(\dfrac{x}{c}+t\right) + F_2\left(\dfrac{x}{c}-t\right)
 \]
 represents two  waves of arbitrary forms running in opposite directions,
 or the Klein-Gordon equation (see \cite{Grim,Grim10,Pelin19,Petr20} and
 references cited). Recently, the set of reference equations has been
 extended \cite{Kap22,Kap23,Petr23} to include the Euler-Darboux-Poisson
 equation,
 \be
 \dfrac{\ptl^2 f}{\ptl t^2} - \dfrac{\ptl^2 f}{\ptl\xi^2} - \dfrac{2m}{\xi}\,
 \dfrac{\ptl f}{\ptl\xi} = 0, \quad \xi = \int\!\dfrac{\dd x}{c(x)},
 \label{EDP}
 \ee
 and some related equations. For $m=0$ Eq.~(\ref{EDP}) is equivalent to
 Eq.~(\ref{WE}), and for $m=1$ it describes spherically symmetric traveling
 waves in homogeneous three-dimensional environment. For any integer $m$,
 its solution also describes two independent traveling waves, each in the
 form of a finite sum (see, for example, \cite{Mises}, \S 12.4),
 \be
 f(\xi,t) = \sum_{k=0}^{|m|}C_{k;m}\xi^{k+m_0}\dfrac{\ptl^k F}{\ptl\xi^k},
 \qquad F(\xi,t) = F_1(\xi+t) + F_2(\xi-t), \quad C_{k;m}=\mbox{const},
 \label{TW}
 \ee
 where $F_1$ and $F_2$ are arbitrary functions, $m_0=1-2m$ for $m>0$, $m_0=0$
 for $m<0$, $C_{0;m}=1$, and the remaining constants can be easily found by
 substituting into Eq.~(\ref{EDP}). We call attention to the fact that
 $F_1(Z)$ serves as a generating function for all functions dependent on the
 phase $(\xi+t)$ of the first wave, and $F_2(Z)$ plays the same role for the
 functions dependent on the phase $(\xi-t)$ of the second wave.

 Taking these results into account, one can formulate an extended concept of
 traveling waves. Namely, such a wave [with the speed $c(x)$] can be
 described by a finite sum,
 \be
 f(x,t) = \sum_{k=1}^{n}a_k(x)\Phi_k\left(t-\!\din\dfrac{\dd x}{c(x)}\right)\,,
 \label{TW1}
 \ee
 in which the form factors $a_k(x)$ are determined by the medium arrangement,
 and the phase functions $\Phi_k(Z)$ have a common generating function
 $\Phi(Z)$. In isotropic media, an oppositely propagating traveling wave is
 represented by a similar sum with the same form factors but its own
 generating function $\tilde\Phi\Bl(t+\int\dd x/c(x)\Br)$.

 If the inhomogeneous medium moves with a velocity $V(x)$, the waves are
 carried away by the flow and propagate with different velocities,
 $V(x)+c(x)$ and $V(x)-c(x)$. In addition, the flow creates a preferred
 direction and violates isotropy. Therefore, to search for non-reflective
 configurations of moving media, it was proposed another approach
 \cite{ChSt22,ChSt22a} based on factorization of wave equations. To do this,
 the function $f(x,t)$ describing wave motion is represented as
 $f(x,t)=a(x){\cal F}(x,t)$ and then the conditions for the flow parameters
 and the dependence of $a(x)$ on these parameters are found under which the
 wave equation takes the form
 \be
 \left[\dfrac{\ptl}{\ptl t} + w_1(x)\,\dfrac{\ptl}{\ptl x} + G(x)\right]
 \left[\dfrac{\ptl}{\ptl t} + w_2(x)\,\dfrac{\ptl}{\ptl x}\right]
 {\cal F}(x,t) = 0,
 \label{Eq}
 \ee
 where $w_{1,2}(x)$ are the wave velocities.

 By this method, wide classes of shallow water flows were found, in which
 surface \cite{ChSt22,ChSt22a} or internal \cite{Chur23,Chur23a,Chur24}
 gravity waves can propagate without reflection. It turned out that for all
 these diverse flows $G(x)$ is equally related to $w_{1,2}(x)$,
 \be
 G(x) = G_0(x) = \dfrac{w_1(x)w'_2(x)-w'_1(x)w_2(x)}{w_1(x)-w_2(x)},
 \label{G0}
 \ee
 and is invariant under the permutation of $w_1$ and $w_2$ (hereinafter the
 prime denotes the derivative with respect to the function argument).
 Moreover, the equation (\ref{Eq}) is invariant as well and the general
 solution to the problem,
 \be
 f(x,t) = a(x){\cal F}(x,t) =
 a(x)\left[{\cal F}_a\left(t - \din_{x_0}^{x}\dfrac{\dd x_1}{w_1(x_1)}
 \right) +
 {\cal F}_b\left(t - \din_{x_0}^{x}\dfrac{\dd x_1}{w_2(x_1)}\right)\right],
 \quad x_0 = \mbox{const},
 \label{Sol0}
 \ee
 is a superposition of two waves of arbitrary shape having a common form
 factor $a(x)$ specified by the flow arrangement.

 A similar study for Alfv\'{e}n waves in plasma flows along magnetic field
 \cite{Chur24a} has shown, however, that there are two types of non-reflective
 currents. For the currents of one type, $G=G_0$ and the wave structure is
 described by Eq.~(\ref{Sol0}). By contrast, for the currents of another type,
 even very similar structurally, $G\ne G_0$, the symmetry with respect to
 $w_1$ and $w_2$ permutation is broken, and the waves differ in both the speed
 and structure.

 Equation (\ref{Eq}) is integrable in quadratures. In its general solution
 \be
 {\cal F}(x,t) = \!\din_{x_0}^{x}\!\dfrac{\dd x_1}{w_2(x_1)}\,\exp\left[\!-
 \!\din_{x_0}^{x_1}\!\dfrac{\dd x_2G(x_2)}{w_1(x_2)}\right]H_1\left(t\!-\!
 \!\din_{x_0}^{x_1}\!\dfrac{\dd x_3}{w_1(x_3)} \!-\! \!\din_{x_1}^{x}\!
 \dfrac{\dd x_3}{w_2(x_3)}\right) + H_2\left(t\!-\!
 \!\din_{x_0}^{x}\!\dfrac{\dd x_1}{w_2(x_1)}\right)
 \label{f}
 \ee
 an arbitrary function $H_2(Z)$ describes the second wave (running at speed
 $w_2(x)$). The argument of another arbitrary function, $H_1(Z)$, is
 arranged in such a manner that a part of the path between $x_0$ and $x$ is
 passed with the speed of the first wave, $w_1(x)$, and the remainder --
 with the speed of the second wave. In general case, the corresponding
 contribution to ${\cal F}(x,t)$ describes a mutual transformation of waves
 in the course of their propagation. In other words, if $H_1=0$, there is
 the second wave only. But if $H_1\ne 0$, then both waves a generally
 present for any $H_2(Z)$. From a physical point of view, such a fundamental
 difference between waves seems strange.

 In the case of Alfv\'{e}n waves, it was possible to show that the solution
 (\ref{f})  with properly chosen $H_2(Z)$ does not contain the second wave,
 hence, there is no wave transformation. This fact is a consequence of the
 specific to Alfv\'{e}n waves dependence of $G$ on the flow parameters. Our
 main objective is to find out what a dependence of $G$ on $w_1$ and $w_2$
 should be in order that Eq.~(\ref{Eq}) will describe a pair of
 independently propagating waves, and what the structure these waves should
 have. Section~\ref{sec2} is devoted to a detailed analysis of the problem.
 The results obtained and their relation to the results of previous studies
 are discussed in Sec.~\ref{sec3}.
 \section{Conditions for an independent propagation of waves}
 \label{sec2}
 \hspace\parindent
 Let us introduce the operators of differentiation along characteristics,
 \[
 \cD_{1,2} = \dfrac{\ptl}{\ptl t} + w_{1,2}(x)\,\dfrac{\ptl}{\ptl x},
 \qquad \cD_1\cD_2 - \cD_2\cD_1 = G_0(x)\Bl(\cD_1 - \cD_2\Br).
 \]
 Then Eq.~(\ref{Eq}) can be written in two equivalent forms,
 \[
 \Bl[\cD_1 + G(x)\Br]\cD_2 {\cal F}(x,t) = 0
 \]
 and
 \be
 \ba{l}
 \Bl[\cD_2 + q_1(x)\Br]\Bl[\cD_1 + q_2(x)\Br]{\cal F}(x,t) -
 k(x){\cal F}(x,t) = 0,
   \\ \\
 q_1(x) = G_0(x), \quad q_2(x) = G(x) - G_0(x), \quad
 k(x) = \Bl[\cD_2 + G_0(x)\Br]\Bl[G(x) - G_0(x)\Br].
 \ea
 \label{Eq1}
 \ee
 If $G(x)=G_0(x)$, $q_2(x)=k(x)=0$ and both
 equations have a factorized form. They differ by the permutation of
 $w_1$ and $w_2$ only and, as noted in Introduction, ${\cal F}(x,t)$
 is given by Eq.~(\ref{Sol0}).

 To study in detail the cases when $G(x)\ne G_0(x)$, we introduce
 $G_1(x)$ by the relation
 \be
 \dfrac{w_1(x)-w_2(x)}{w_2(x)}\,G_1(x) = G(x) - G_0(x),
 \label{G}
 \ee
 represent the general solution of Eqs.~(\ref{Eq}) and (\ref{Eq1}) in the
 form
 \be
 {\cal F}(x,t) = {\cal F}_1(x,t) + {\cal F}_2\left(t\!-\!
 \!\din_{x_0}^{x}\!\dfrac{\dd x_1}{w_2(x_1)}\right)
 \label{F12}
 \ee
 with arbitrary ${\cal F}_2(Z)$, and then try to find such $G_1(x)$, for
 which ${\cal F}_1(x,t)$ describes a traveling wave with velocity
 $w_1(x)$, that is, a wave with the structure similar to Eq.~(\ref{TW1}),
 \be
 {\cal F}_1(x,t) = \sum_{m=1}^{n}a_m(x)F_m(T), \quad
 T = t -\!\din_{x_0}^{x}\!\dfrac{\dd x_1}{w_1(x_1)}\,.
 \label{F1}
 \ee

 The Laplace cascade method for integrating hyperbolic equations improved
 by Legendre and Imshenetsky (see \cite{Imsh,Goursat} and, for a more
 modern presentation, \cite{Baikov,Kap24}) will help us with this. It
 consists in the step-by-step replacement of the dependent variable,
 \[
 {\cal F}_{1,i-1}(x,t) = \dfrac{\cD_2 + q_{1,i-1}(x)}{k_{i-1}(x)}\,
 {\cal F}_{1,i}(x,t),\ \ {\cal F}_{1,0} = {\cal F}_1,\ \ k_0 = k,\ \
 q_{1,0} = q_1,\ \ i = 1,\,2,\dots,\,n,
 \]
 which transforms Eq.~(\ref{Eq1}) into equations of a similar form,
 \[
 \Bl[\cD_2 + q_{1,i}(x)\Br]\Bl[\cD_1 + q_{2,i}(x)\Br]{\cal F}_{1,i}(x, t)
 - k_i(x){\cal F}_{1,i}(x,t) = 0.
 \]
 For our study, the key result of the method is that ${\cal F}_1(x,t)$ can
 be represented in the form (\ref{F1}) if and only if there exists such $n$
 that $k_n(x)=0$. In that case, the corresponding equation is factorized and
 all the functions $F_m(T)$ are expressed in terms of derivatives of an
 (arbitrary) function $F_1(T)$, so that (cf. Eq.~(\ref{TW}))
 \be
 {\cal F}_1(x,t) = a_1(x)F_1(T) + \sum_{m=1}^{n-1}a_{m+1}(x)F_1^{(m)}(T),
 \quad F_1^{(m)}(T) = \dfrac{\dd^{m}F_1(T)}{\dd T^m}.
 \label{F1a}
 \ee

 Substituting the expansion (\ref{F1a}) into Eq.~(\ref{Eq}) yields
 \be
 \ba{l}
 F_1(T)\,\hL\Bl[w_2(x)\,a'_1(x)\Br] + \sum\limits_{m=1}^{n-1}F_1^{(m)}( T)
 \,\hL \left[\dfrac{w_1(x)-w_2(x)}{w_1(x)}\,a_m(x) + w_2(x)\,a'_{m+1}(x)
 \right]
   \\ \\ \phantom{WWW}
 + F_1^{(n)}(T)\,\hL\left[\dfrac{w_1(x)-w_2(x)}{w_1(x)}\,a_n(x)\right]
 = 0, \qquad \hL =\dfrac{\dd}{\dd x} + \dfrac{G(x)}{w_1(x)}\,.
 \ea
 \label{EqFa}
 \ee
 Assuming that $w_{1,2}(x)$ are continuous and $w_1(x)\ne w_2(x)$ everywhere,
 we introduce new variable
 \be
 \vp(x) = \!\din_{x_0}^{x}\!\dd x_1\left[\dfrac{1}{w_1(x)}-\dfrac{1}{w_2(x)}
 \right]
 \label{fi}
 \ee
 monotonically varying with $x$. In what follows, it is assumed that $w_{1,2}$,
 $G$, $G_1$, and $a_m$ depend on $\vp$, and the prime will continue in
 use for derivative with respect to the argument ($T$ or $\vp$, but not $x$).
 One can easily seen that for an arbitrary function $f(x)$
 \[
 w_2\dfrac{\dd f}{\dd x} = -\dfrac{w_1 - w_2}{w_1}\,\dfrac{\dd f}{\dd\vp},
 \quad \hL\left(\dfrac{w_1 - w_2}{w_1}\,f\right) =
 -\dfrac{[w_1 - w_2]^2}{w_1^2w_2}\,\hL_1f, \quad
 \hL_1 = \dfrac{\dd}{\dd\vp} - G_1,
 \]
 so that Eq.~(\ref{EqFa}) simplifies significantly:
 \be
 F_1(T)\,\hL_1a'_1(\vp) + \sum_{m=1}^{n-1}F_1^{(m)}(T)\,\hL_1\Bl[a'_{m+1}(\vp)
 - a_m(\vp)\Br] - F_1^{(n)}(T)\,\hL_1a_n(\vp) = 0.
 \label{EqFa1}
 \ee

 Since $F_1(T)$ is arbitrary, the coefficients at it and its derivatives must
 vanish, therefore
 \be
 \hL_1a'_1(\vp) = \hL_1a_n(\vp) = 0\ \ \mbox{and}\ \ \hL_1\Bl[a'_{m+1} -
 a_m(\vp)\Br] = 0, \quad 1 \le m \le n-1.
 \label{mn}
 \ee
 To find solutions to these equations, it is convenient to introduce the
 functions
 \be
 \ba{l}
 E_{m+1}(\vp) = \!\din_{0}^{\vp}\!E_m(\vp_1)\dd\vp_1,\ \ m = 0,\,1,\,2,\dots;
   \\ \\
 E_0(\vp) \equiv E(\vp) = \exp\left[\din_{0}^{\vp}\!G_1(\vp_1)\dd\vp_1
 \right], \quad \hL_1E(\vp) = 0.
 \label{E}
 \ea
 \ee
 Without loss of generality, we may put $a_n(\vp)=E(\vp)$ and
 $a'_1(\vp)=\gm E(\vp)$, where $\gm=\mbox{const}$. Remaining
 Eqs.~(\ref{mn}) can be easily integrated as well:
 \[
 \ba{l}
 a'_1(\vp) = \gm E(\vp)\ \ \Longrightarrow\ \ a_1(\vp)= \gm E_1(\vp)+\beta_1,
    \\ \\
 a'_2(\vp) = a_1(\vp)+2\al_1E(\vp)\ \ \Longrightarrow\ \
% \\ \\ \phantom{wwww}
 a_2(\vp) = \gm E_2(\vp) + 2\al_1E_1(\vp) + \beta_1\vp + \beta_2,\dots,
    \\ \\
 a_{m+1}(\vp) = \gm E_{m+1}(\vp) + 2\al_1E_m(\vp) + \dots + 2\al_mE_1(\vp)+
 \beta_1\dfrac{\vp^m}{m!} + \dots + \beta_m\vp + \beta_{m+1},
 \ea
 \]
 where $\al_i$ and $\beta_i$ are constants, $m<n$. For $m=n-1$ we obtain the
 equation
 \be
 a_n(\vp) = \gm E_n(\vp) + 2\al_1E_{n-1}(\vp) + \dots + 2\al_{n-1}E_1(\vp)+
 \beta_1\dfrac{\vp^{n-1}}{(n-1)!} + \dots + \beta_n = E(\vp).
 \label{an}
 \ee

 Denoting $y(\vp)=E_n(\vp)$ and taking into account that (see Eq.~(\ref{E}))
 $y^{(i)}(\vp)=E_{n-i}(\vp)$, one can re-write Eq.~(\ref{an}) in the form of
 a $n$-th order differential equation
 \be
 y^{(n)} - 2\al_{n-1}y^{(n-1)} - \dots - 2\al_1y' - \gm y =
 \beta_1\dfrac{\vp^{n-1}}{(n-1)!} + \dots + \beta_{n-1}\vp + \beta_n
 \label{yn}
 \ee
 with initial conditions following from Eqs.~(\ref{E}),
 \be
 y(0) = y'(0) = \dots = y^{(n-1)}(0) = 0, \quad y^{(n)}(0) = 1.
 \label{InC}
 \ee
 The last condition means that $\beta_n=1$.

 For $\gm\ne 0$, the solution to Eq.~(\ref{yn}) has the form
 \be
 y(\vp) = \sum_{m=1}^{n}D_m\exp(\lb_m\vp) + P_n(\vp), \quad
 P_n(\vp) = \sum_{k=1}^{n-1}B_k\,\dfrac{\vp^k}{k!}\,.
 \label{yn1}
 \ee
 Here the polynomial $P_n(\vp)$ is a particular solution to the
 inhomogeneous equation, $B_{n-1}=-\beta_1/\gm$, other constants $B_k$ are
 calculated by substituting into Eq.~(\ref{yn}), the constants $D_m$ can be
 found from the initial conditions (\ref{InC}), and $\lb_m$ are the roots
 of the characteristic equation
 \[
 \lb^n - 2\al_{n-1}\lb^{n-1} - \dots - 2\al_1\lb - \gm = 0.
 \]
 This solution, like Eq.~(\ref{yn}) itself, depends on $(2n-1)$ arbitrary
 constants $\gm$, $\al_i$ and $\beta_i$\ \ $(i=1,\dots,n-1)$.
 Finally, employing the relations
 \be
 E(\vp) = y^{(n)}(\vp), \qquad G_{1;n}(\vp) \equiv \dfrac{E'(\vp)}{E(\vp)}
 = \dfrac{y^{(n+1)}(\vp)}{y^{(n)}(\vp)}\,
 \label{G1}
 \ee
 and taking the wave velocities $w_1$ and $w_2$ as given, we find a family
 of functions $G_{1;n}(\vp)$ depending on $(2n-1)$ parameters, for which
 both waves propagate independently. At last, it should be noted that
 if $\gm=\al_1=\dots=\al_p=0$, but $\al_{p+1}\ne 0$ ($p=0,\,1,\,2,\dots$),
 the upper limit of summation in $P_n(\vp)$ increases to $(n+p)$ and
 $B_{n+p}=-\beta_1/(2\al_{p+1})$.

 Let us now consider some special cases. For $n=1$, Eqs.~(\ref{yn}) and
 (\ref{InC}) give
 \[
 y(\vp) = E_1(\vp), \quad y' - \gm y = 1,\quad y(0) = 0\ \
 \Longrightarrow\ \ y(\vp) = \dfrac{\exp(\gm\vp)-1}{\gm}\,
 \]
 whence it follows that
 \be
 a_1(\vp) = E(\vp) = \exp(\gm\vp), \qquad G_{1;1}(\vp) = \gm = \mbox{const}.
 \label{n1}
 \ee
 If $\gm=0$, then $G_1=0$, $G=G_0$, $a_1=1$, and the waves have a similar
 structure (see Eq.~(\ref{Sol0})).

 For $n=2$, from Eqs.~(\ref{yn}) and (\ref{InC}) we find
 \[
 \ba{l}
 y(\vp) = E_2(\vp),\ \ y'' - 2\al_1y' - \gm y = \beta_1\vp + 1,\ \
 y(0) = y'(0) = 0;\\ \\
 P_2(\vp) = -\dfrac{\beta_1}{\gm}\,\vp +
 \dfrac{2\al_1\beta_1-\gm}{\gm^2}\,.
 \ea
 \]
 The roots of the characteristic equation,
 \[
 \lb^2 - 2\al_1\lb - \gm = 0\ \ \Longrightarrow\ \ \lb_{1,2} = \al_1
 \pm\sqrt{\al_1^2+\gm},
 \]
 can be either real or complex.

 Let us begin with real roots, $\al_1^2+\gm > 0$. The solution obeying
 initial conditions can be written in the form
 \[
 \ba{l}
 y(\vp) = \Bl[C_1\cosh(\sg\vp) + C_2\sinh(\sg\vp)\Br]\exp(\al_1\vp) +
 P_2(\vp),
   \\ \\
 C_1 = -\dfrac{2\al_1\beta_1-\gm}{\gm^2}\, \quad
 C_2 = \dfrac{1}{\sg}\left(\dfrac{\beta_1}{\gm} - \al_1D_1\right),
 \quad \sg=|\al_1^2+\gm|^{1/2}.
 \ea
 \]
 Using Eqs.~(\ref{G1}) we find
 \be
 \ba{l}
 a_1(\vp) = \gm y' + \beta_1 = \left[\beta_1\cosh(\sg\vp) -
 \dfrac{\al_1\beta_1-\gm}{\sg} \,\sinh(\sg\vp)\right]\exp(\al_1\vp),
   \\ \\
 a_2(\vp) = E(\vp) = \left[\cosh(\sg\vp) +\dfrac{\al_1+\beta_1}{\sg}\,
 \sinh(\sg\vp)\right]\exp(\al_1\vp),
 \quad G_{1;2}(\vp) = \dfrac{E'(\vp)}{E(\vp)}\,.
 \ea
 \label{n2p}
 \ee
 When $\al_1^2+\gm=0$ or $\al_1^2+\gm < 0$, the solution is constructed in
 a similar way and can be obtained from Eqs.~(\ref{n2p}) by passing to the
 limit $\sg\to 0$ or replacing $\sg$ by $i\sg$, respectively. As a result,
 \be
 \ba{l}
 \gm = -\al_1^2\ \ (\sg=0): \qquad
 a_1(\vp) = \Bl[\beta_1 - \al_1(\al_1+\beta_1)\vp\Br]\exp(\al_1\vp),
   \\ \\
 a_2(\vp) = E(\vp) = \Bl[1 + (\al_1+\beta_1)\vp\Br]\exp(\al_1\vp),
 \quad G_{1;2}(\vp) = \al_1 + \dfrac{\al_1+\beta_1}{1 + (\al_1+\beta_1)\vp}\,,
 \ea
 \label{n20}
 \ee
 and
 \be
 \ba{l}
 \al_1^2+\gm < 0: \qquad
 a_1(\vp) = \left[\beta_1\cos(\sg\vp) - \dfrac{\al_1\beta_1-\gm}{\sg}
 \,\sin(\sg\vp)\right]\exp(\al_1\vp),
   \\ \\
 a_2(\vp) = E(\vp) = \left[\cos(\sg\vp) +
 \dfrac{\al_1+\beta_1}{\sg}\,\sin(\sg\vp)\right]\exp(\al_1\vp),
 \quad G_{1;2}(\vp) = \dfrac{E'(\vp)}{E(\vp)}\,.
 \ea
 \label{n2m}
 \ee
 \section{Discussion}
 \label{sec3}
 \hspace\parindent
 The above analysis of solutions of Eq.~(\ref{Eq}) has shown that, for
 given wave velocities $w_1(x)\ne w_2(x)$ in a stationary inhomogeneous
 medium, there is a hierarchy of $(2n-1)$-parametric families of
 functions $G_{1;n}(x)$ for which both waves are traveling. The problem
 of calculating the $n$-th family is reduced to solving the $n$-th order
 linear differential equation (\ref{yn}) with constant coefficients and
 initial (at $x=x_0$, that is, $\vp=0$) conditions (\ref{InC}).

 The families of functions $G_{1;n}$ are embedded in each other in the
 sense that the family $G_{1;n}$ is a subset of the family $G_{1;n+1}$ and
 can be selected by specifying two relationships between parameters. So,
 if we take particular cases considered above, we see that $G_{1;2}(\vp)$
 converts (up to notation) into $G_{1;1}(\vp)$ when $\gm=-\al_1^2$ and
 $\beta_1=-\al_1$ (compare Eqs.~(\ref{n1}) and (\ref{n20})).

 Let us pay attention to the fact that the structure of the second wave
 is the same in all cases, whereas the structure of the first one
 depends significantly on $n$ (see Eqs.~(\ref{F12}) and (\ref{F1})).
 This is a consequence of the problem statement because we start with the
 already factorized Eq.~(\ref{Eq}). In the general case of a non-factorized
 wave equation, %but describes two traveling waves,
 its traveling wave solutions should be searched for in the form (see
 \cite{Goursat,Kap24})
 \[
 \ba{l}
 {\cal F}(x,t) = a(x)F_1(Z_1) + \sum\limits_{i=1}^{n-1}a_i(x)F_1^{(i)}(Z_1) +
 b(x)F_2(Z_2) + \sum\limits_{i=1}^{m-1}b_i(x)F_2^{(i)}(Z_2), \\ \\
 Z_{1,2} = t - \din_{x_0}^{x}\!\dfrac{\dd x_1}{w_{1,2}(x_1)}\,,
 \ea
 \]
 with arbitrary $F_1(Z_1)$ and $F_2(Z_2)$, and, generally, $m\ne n$. However,
 corresponding calculation of $G_1(x)$ appears to be much more cumbersome.

 When applying the results obtained, it should be borne in mind that the
 physical statement of the problem dictates a very definite relation
 between $G_1(x)$ and wave velocities $w_{1,2}(x)$ (and, possibly, other
 parameters of the medium). Therefore, identifying such $G_1$ with some
 $G_{1;n}$ calculated above means {\it the imposing of restrictions on
 the flow parameters} which select non-reflective flows from all
 physically possible ones.

 In conclusion, let us briefly dwell on previously studied non-reflective
 flows. The problems of propagation without reflection of surface
 \cite{ChSt22,ChSt22a} and internal \cite{Chur23,Chur23a,Chur24} gravity
 waves in shallow water flows, as well as a part of plasma flows supporting
 traveling Alfv\'{e}n waves (see \cite{Chur24a}, Sec.~III), correspond to the
 case $n=1$ and $\gm=0$, with equally structured waves (see Eq.~(\ref{Sol0})).
 The other part of the non-reflective plasma flows corresponds to $n=2$ and,
 therefore, to waves of different structure. Namely, the first wave is
 described by Eq.~(\ref{n2p}) or (\ref{n2m}) with $\al_1=0$ (see
 \cite{Chur24a}, Sec.~IV). It is interesting to note that the seemingly
 reasonable case $n=1$, $\gm\ne 0$, when the waves have a similar structure
 but different form factors, has been encountered in none of the physical
 problems yet considered.

 \end{document}